\documentclass{article}
\usepackage{bbold}
\usepackage{url}
\usepackage[margin=1in]{geometry}
\usepackage[utf8]{inputenc}
\usepackage{amsmath}
\usepackage{bm}
\usepackage{graphicx}
\usepackage{float}
\usepackage{setspace}
\usepackage{color}
\usepackage{soul}
\usepackage{subfigure}
\usepackage{subcaption}
\usepackage{multirow}
\usepackage{booktabs}

\usepackage[left]{lineno}

\title{Improving Outbreak Forecasts Through Model Augmentation}
\author{Graham C. Gibson$^{1^*}$, Spencer J. Fox$^{3^*}$, Emily Javan$^2$, Susan E. Ptak$^2$, Oluwasegun M. Ibrahim$^2$, \\ Michael Lachmann$^3$, Lauren Ancel Meyers$^{2,4}$}

\date{%
    $^1$Los Alamos National Laboratory, Los Alamos, NM\\
    $^2$University of Texas at Austin, Austin, TX\\
    $^3$ University of Georgia, Athens, GA\\
    $^4$Santa Fe Institute, Santa Fe, NM\\
    $^*$These authors contributed equally\\[2ex]%
    \today
}
\begin{document}
\maketitle

\section*{Abstract}
Accurate forecasts of disease outbreaks are critical for effective public health responses, management of healthcare surge capacity, and communication of public risk. There are a growing number of powerful forecasting methods that fall into two broad categories—empirical models that extrapolate from historical data, and mechanistic models based on fixed epidemiological assumptions. However, these methods often underperform precisely when reliable predictions are most urgently needed—during periods of rapid epidemic escalation. Here, we introduce \textit{epimodulation}, a hybrid approach that integrates fundamental epidemiological principles into existing predictive models to enhance forecasting accuracy, especially around epidemic peaks. When applied to simple empirical forecasting methods (ARIMA, Holt-Winters, and spline models), epimodulation improved overall prediction accuracy by an average of 9.1\% (range: 8.2–12.5\%) for COVID-19 hospital admissions and by 19.5\% (range: 17.6–23.2\%) for influenza hospital admissions; accuracy during epidemic peaks improved even further, by an average of 20.7\% and 25.4\%, respectively. Epimodulation also substantially enhanced the performance of complex forecasting methods, including the COVID-19 Forecast Hub ensemble model, demonstrating its broad utility in improving forecast reliability at critical moments in disease outbreaks.

\section*{Significance Statement}
Reliable outbreak forecasting is essential for public health decision-making, yet traditional methods often falter during critical epidemiological time periods such as the peak of an epidemic. This work introduces \textit{epimodulation}, a novel forecasting approach that integrates epidemiological principles into a wide-range of forecasting models to enhance their performance. The approach increases model accuracy by up to 25.4\% during epidemic peaks for COVID-19 and influenza hospital admission forecasts, without reducing accuracy at other times. Due to its flexible and straightforward design, epimodulation is a powerful tool for enhancing forecast performance, boosting healthcare preparedness and risk communication during critical outbreak periods.

\section*{Introduction}
Forecasting epidemiological trends, such as cases, hospital admissions, or deaths, can provide vital situational awareness for decision makers and communities \cite{lutz_applying_2019}. During the COVID-19 pandemic, forecasts informed the implementation and relaxation of social distancing and face mask policies, the submission of timely healthcare staffing requests, and the construction of alternate care sites to manage healthcare overflow \cite{fox_real-time_2022,biggerstaff_improving_2022}. However, many COVID-19 forecasting models struggled to accurately predict the timing and magnitude of peaks, exactly when they would have been most useful for communicating changing risks and managing limited healthcare resources \cite{reich_predictability_2021, biggerstaff_improving_2022, cramer_evaluation_2022,castro_turning_2020,case_accurately_2023}.

First quantified in the 1800s, the shape of an epidemiological curve is a cornerstone of infectious disease epidemiology \cite{Brownlee250}. Epidemics typically follow a pattern of early exponential growth, peaking at a maximum value, and then declining as either pharmaceutical or nonpharmaceutical interventions take hold, behaviors change to reduce transmission, or the supply of susceptible individuals decreases through recovery and immunity. While curve shape varies by disease and context, all epidemics share this basic trajectory. Epidemic forecasts aim to predict key features of the curve: when the epidemic will begin, how quickly it will grow, the timing and height of the peak, the rate of decline, and whether it will fully extinguish. Among these, forecasting the peak is often the most critical for managing risks and allocating resources \cite{lutz_applying_2019,fox_real-time_2022,biggerstaff_improving_2022}. In addition to the public health considerations, absolute performance metrics, such as mean absolute error (MAE) and weighted interval score (WIS), reward models that accurately capture the peak by placing greater emphasis on periods with the highest values \cite{gneiting_strictly_2007,bracher_evaluating_2021}.

The COVID-19 Forecasting Hub, established in 2020, collected, ensembled, and evaluated over 92 million predictions from 110 unique models between April 2020 and May 2022 \cite{covidhub}. Forecasting techniques broadly fall into two categories: \textit{mechanistic} models, which explicitly incorporate known epidemiological processes, and \textit{empirical} models, which identify patterns without making such assumptions \cite{biggerstaff_results_2016, mcgowan_collaborative_2019,reich_collaborative_2019}. Mechanistic models often rely on the canonical susceptible-infected-recovered (SIR) framework, network, or agent-based extensions \cite{he2020seir, pandey2020seir, gibson2020real, piccolomini2020preliminary, pei2020differential, VENKATRAMANAN201843, fox_real-time_2022}, while empirical models draw on a variety of time-series, non-parametric density estimation methods, and artificial intelligence or machine learning  methods \cite{barria2021prediction,ray2017infectious, chimmula_time_2020, assad_comparing_2023, kumar_covid-19_2020,sesti2021integrating, venkatramanan_forecasting_2021}. These approaches have produced high-performing models \cite{lutz_applying_2019,biggerstaff_results_2016,mcgowan_collaborative_2019,reich_collaborative_2019}, with empirical often excelling in direct comparisons \cite{reich_collaborative_2019,kandula2018evaluation, mcgowan2019collaborative, baker2018mechanistic}. Mechanistic models, however, may hold an advantage in forecasting peaks, especially for novel threats with limited historical data, as they explicitly account for the depletion of susceptible individuals in the population. Most empirical models do not encode the characteristic shape of an epidemic \cite{chimmula_time_2020,assad_comparing_2023,kumar_covid-19_2020}, though some use shape-based models to capture seasonal trends \cite{brooks_flexible_2015,noauthor_xxiv_1825,woody_projections_2020}. 

Here, we introduce a model augmentation method called ``epimodulation" which integrates epidemiological dynamics into any forecast model. In essence, epimodulation encodes the susceptible depletion process into simple empirical forecast models with a single additional parameter. We show that augmented models quickly learn epidemiological dynamics and consistently outperform their base versions, particularly during epidemic peaks. Applying epimodulation to ensemble forecasts likewise improves performance. These findings suggest a low-dimensional approach to improving a wide range of infectious disease forecast models.

\section*{Methods}
\subsection*{Encoding epidemiological peak behavior}

We start with a traditional differential equation-based compartmental model of infectious disease transmission and then re-express the equations to provide an epimodulated statistical model that takes into account epidemic peaks \cite{kermack1927contribution,miller2017mathematical}.

We begin with the SIR model structure from Miller et al. \cite{miller2017mathematical}.

\begin{eqnarray}
    S(t) &= S_0 e^{-\beta\int_0^t I(t')dt'}\\
    I(t) &= 1 - R(t) - S(t)\\
    R(t) &= R_0 + \gamma \int_0^{t} I(t')dt' 
\end{eqnarray}

with the symbol definitions given in Table \ref{table:notation}.

\begin{table}[H]
\centering
\caption{Notation Summary}
\begin{tabular}{ll}
\toprule
Symbol & Description \\
\midrule
$S(t)$ & Susceptible proportion at time $t$ \\
$S_0$ & Initial susceptible proportion\\
$I(t)$ & Infected proportion at time $t$ \\
$R(t)$ & Recovered proportion at time $t$ \\
$\beta$ & Transmission rate \\
$\gamma$ & Recovery rate \\
$\theta$ & Epimodulation parameter \\
$\phi$ & Initial condition on susceptibles \\
$F(t)$ & Latent transmission function \\
\label{table:notation}
\end{tabular}
\end{table}

We derive the expression for the incidence of the infection, $i(t)$. This variable allows us to relate the model to the observed data, assuming that reported new cases and hospital admissions are proportional to the total number of new infections. 

\begin{eqnarray}
    i(t)&=& -\frac{d}{dt}S(t) \\
    i(t)&=& \beta S_0 I(t) e^{-\beta\int_0^t I(t')dt'} 
\end{eqnarray}

Equation (5) can be restated as given by 
\begin{equation}
   i(t;\phi,\theta)= \underbrace{\phi \theta F(t)}_{\scriptsize \text{Infection}} \hspace{.25cm} \underbrace{e^{-\theta \int_0^{t} F(t')dt'}}_{\raisebox{-0.75ex}{\scriptsize \text{Susceptible Depletion}}}
\end{equation}

where $\phi$ is an initial proportion of susceptibles, and $\theta$ is the rate of susceptible depletion as the epidemic progresses. This equation reduces to the original SIR model when $F(t) = I(t)$, $\phi = S_0$, and $\theta = \beta$. The Gompertz curve is also a special case of this function. 

If we examine Equation 6, the only piece relevant to susceptible depletion is the term $e^{-\theta \int_0^{t} I(t')dt'}$. This term is bounded between [0,1] and multiplies $S_0$, essentially scaling down the initial number of susceptibles multiplicatively as the epidemic progresses. If this term is $1$ then all susceptibles remain and we have $i(t) = \beta S_0 I(t)$ implying all currently infected individuals cause an infection with probability $\beta$, if this term is $0$ then there are no new infections. We use this intuition to modify general statistical models as follows. Suppose we have an arbitrary statistical model $\mathcal{M}$ capable of making forecasts for times $t:(t+k)$ as given by $\hat{y}_{t:(t+k)}$. Epimodulation integrates susceptible depletion into the forecast as given by,

\begin{equation}
    \tilde{y}_{t:(t+k)}=\hat{y}_{t:(t+k)} e^{-\theta \sum_{i=t}^{t+k} \hat{y}_{i}}
\end{equation}

That is, we take the forecasts and apply the $[0,1]$ scaling based on Equation 6. This prevents forecasts from increasing indefinitely without incorporating any explicit peak dynamics into the underlying model. The term $e^{-\theta \sum_{i=t}^{t+k} \hat{y}_i}$ increases proportional to the magnitude of $\hat{y}$. That is, larger forecasted burden metrics lead to larger epimodulation. Note that the peak does not necessarily occur within the forecast window (Figure 1), however, it guarantees that the model will eventually peak. The key question then becomes how to estimate $\theta$.

\subsection*{Statistical inference of peak dynamics}

We use cross-validation to estimate $\theta$ in Equation (8) based on the performance of historical forecasts. Specifically, if we condition on the observed data  ($y_1, y_2, \ldots, y_T$) up until time $T$, we can make a $k$ step ahead forecast for all times $t^{*} \in [1, T-k]$ by estimating $\hat{y}_{t^* + 1 : t^* + k}$ (where $1:k$ indicates times 1 through $k$) under $F(t)$ and computing the prediction error: 

\begin{equation}
    PE_{F(t)} = \sum_{t^*=1}^{T-k} (\hat{y}_{t^* + 1 : t^* + k} - {y}_{t^* + 1 : t^* + k})^2
    \label{eq:prediction-error}
\end{equation}

Equation \ref{eq:prediction-error} gives us an estimate of the prediction error if there were no epidemiological dynamics in the forecast model, so we modify $\hat{y}_t^*$ to include those dynamics, leading to an estimate, $PE^{adj}_{\hat{y}_t^*}$, for all forecast dates of: 

\begin{equation}
    PE^{adj}_{F(T)} = \sum_{t^*} (\hat{y}_{t^* + 1 : t^* + k}e^{-\theta \hat{y}_{t^* + 1 : t^* + k}} - {y}_{t^* + 1 : t^* + k})^2
\end{equation}

To estimate $\theta$ in that framework, we carry out an optimization procedure to reduce the prediction error of the epidemiological dynamic forecast model as:

\begin{equation}
    \hat{\theta} =  \text{argmin}_{\theta} \left[ PE^{adj}_{F(T)}(\theta) \right]
\end{equation}

For each retrospective forecast date, we perform the optimization using the ``optim" package in R \cite{R}. 

\subsection*{Retrospective forecasting for COVID-19 and influenza hospital admissions}
To assess the improvement afforded by epimodulation, we retrospectively evaluate three commonly used statistical forecasting models: an automated Autoregressive Integrated Moving Average (ARIMA) model  (as implemented in the auto.arima function in the ``forecast" package), a holt-winters model (as implemented in the ``forecast" package), and a spline model (as implemented in the ``mgcv" package) \cite{forecast,mgcv}. We chose these models because they do not explicitly account for peak dynamics, represent three common and distinct approaches to statistical modeling of infectious disease time series, and they have performed reasonably well in previous forecasting challenges \cite{chimmula_time_2020,assad_comparing_2023,kumar_covid-19_2020,krymova2022trend}. We compared the performance of each models with and without epimodulation.

Each of the six model variations is trained using all data up to a specific forecast date for a specific region, and information is not shared across regions, though sharing information across regions is possible in this framework through a hierarchical model on $\theta$. We chose approaches that do not need predictor covariate data for making forecasts, as the relationship between covariates and epidemiological dynamics can fluctuate during an ongoing outbreak \cite{fox_real-time_2022, nouvellet_reduction_2021,mcdonald_can_2021}, and many potential predictor variables such as mobility or mask wearing estimates were discontinued in early 2022 \cite{apple_covid19_nodate,google_covid-19_nodate,salomon_us_2021}.

To obtain the epidemiological data for retrospective forecasts we used the ``covidHubUtils"  package in R \cite{covidHubUtils}. We obtained daily COVID-19 hospital admission counts for all 50 states, Puerto Rico, Virgin Islands, and Guam from July 14, 2020 through September 11, 2023 and made and evaluated weekly forecasts for 1-28 days from October 12, 2020 to September 11, 2023 following the submission dates and guidelines of the COVID-19 Forecast Hub \cite{cramer_evaluation_2022, covidforecasthub}. We obtained weekly influenza hospital admissions from March 3, 2021 until May 15, 2023, and we made weekly 1-4 week ahead forecasts for influenza hospital admissions from January 10, 2022 until May 15, 2023 following the forecast dates and guidelines of the FluSight forecast hub \cite{flusight1,flusight2}.  We evaluated forecasts using their mean absolute error (MAE), which is a proper scoring criteria for point forecasts \cite{bracher_evaluating_2021}.

\subsection*{Real-time forecasts}
Real-time forecasting in collaborative forecast hubs is the gold standard for proper evaluation of forecast model performance. We tested the performance of our epimodulation procedure on improving the gold standard ensemble performance. For this we extracted probabilistic forecasts from the``COVIDhub-4-week-ensemble model" \cite{covidforecasthub}. Forecasts were submitted in the form of quantiles $\mathcal{Q}$ in order to compute the weighted interval score \cite{bracher2021evaluating}. For each quantile level forecast we applied epimodulation to obtain an epimodulated forecast $\tilde{y}^q_{t:(t+k)}$. 

\begin{equation}
\text{WIS} = \frac{1}{K+1} \left( | y - \hat{y}_m | + \sum_{k=1}^{K} \alpha_k \cdot \text{IntervalScore}_{\alpha_k} \right)
\end{equation}

Where the Interval Score is computed as:

\begin{equation}
\text{IntervalScore}_{\alpha_k} = 
\begin{cases} 
(u_k - l_k) + \frac{2}{\alpha_k} (l_k - y), & \text{if } y < l_k \\
(u_k - l_k), & \text{if } l_k \leq y \leq u_k \\
(u_k - l_k) + \frac{2}{\alpha_k} (y - u_k), & \text{if } y > u_k 
\end{cases}
\end{equation}

We applied the quantile epimodulation across all 50 states from June 7, 2021 until May 29, 2023.

\subsection*{Forecast evaluation}
The COVID-19 Forecast Hub evaluates forecasts using two primary metrics: Mean Absolute Error (MAE) and Weighted Interval Score (WIS). These metrics measure forecast accuracy and uncertainty, balancing point predictions and probabilistic intervals.

MAE quantifies the average magnitude of forecast errors, providing a simple and interpretable measure of prediction accuracy. It is computed as:

\[
\text{MAE} = \frac{1}{N} \sum_{i=1}^{N} \left| \hat{y}_i - y_i \right|
\]

Lower MAE values indicate more accurate point forecasts.

The Weighted Interval Score (WIS) assesses the quality of prediction intervals by penalizing forecasts that are too wide or fail to include the observed value. For a prediction interval \([L_{\alpha}, U_{\alpha}]\) with nominal coverage \(1 - \alpha\), the interval score was calculated as:

\[
\text{IS}_{\alpha} = (U_{\alpha} - L_{\alpha}) + \frac{2}{\alpha} (L_{\alpha} - y) \mathbb{I}(y < L_{\alpha}) + \frac{2}{\alpha} (y - U_{\alpha}) \mathbb{I}(y > U_{\alpha})
\]

where:
\begin{itemize}
    \item \( L_{\alpha} \): Lower bound of the prediction interval
    \item \( U_{\alpha} \): Upper bound of the prediction interval
    \item \( y \): Observed value
    \item \( \mathbb{I}(\cdot) \): Indicator function, equal to 1 if the condition is true and 0 otherwise
\end{itemize}

This metric rewards forecasts with narrow intervals that contain the observed value while penalizing intervals that miss the true outcome or are excessively wide.

The percent improvement in MAE is computed as a relative reduction in MAE compared to the base model forecast. It was defined as:

\[
\text{Percent Improvement} = \frac{\text{MAE}_{\text{base}} - \text{MAE}_{\text{model}}}{\text{MAE}_{\text{base}}} \times 100\%
\]

Higher percent improvement values indicate better performance relative to the baseline. We apply the same equation for percent improvement in WIS.

\begin{figure}
    \centering
    \includegraphics[width=0.8\textwidth]{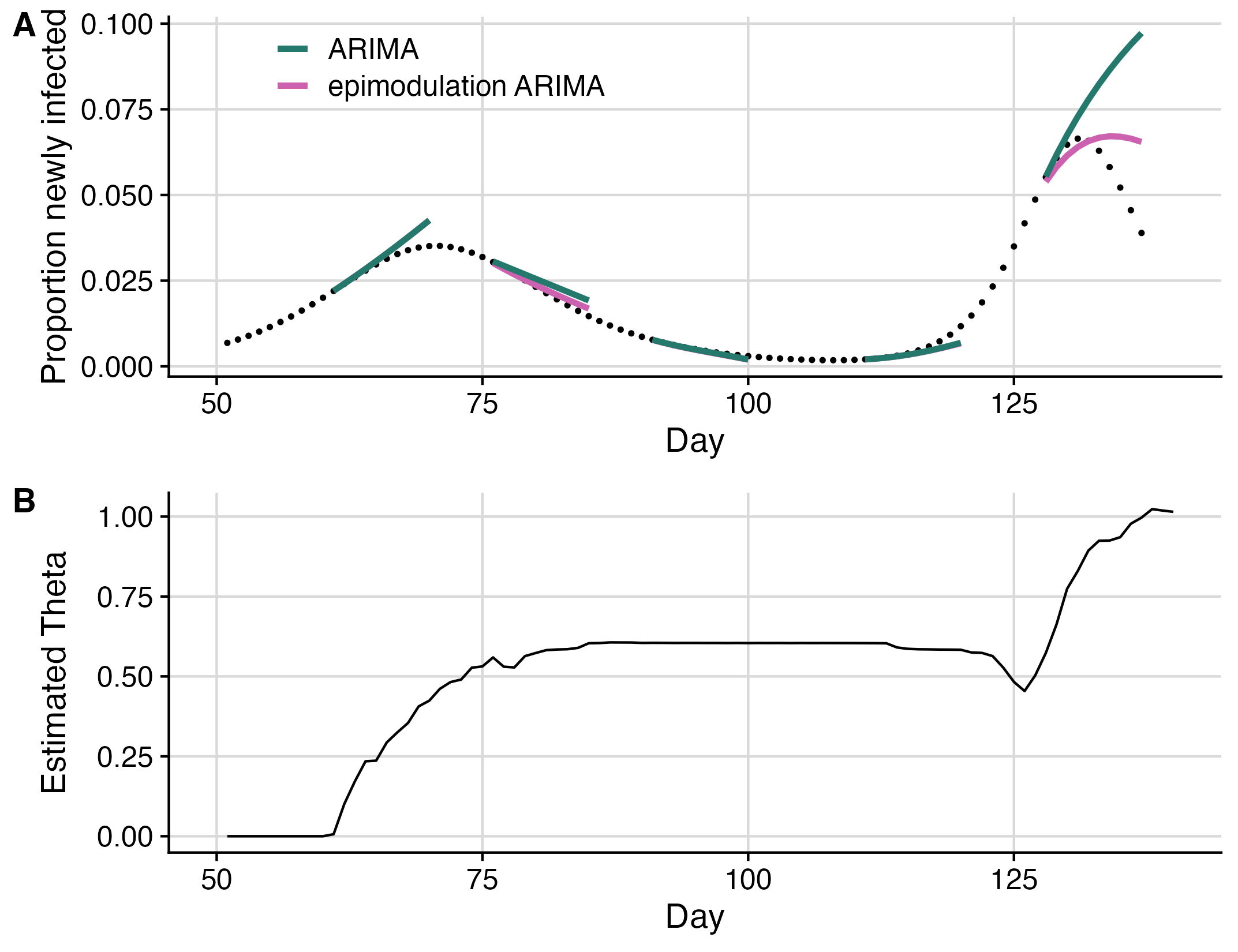}
    \caption{Epimodulated ARIMA model learns peak structure and improves forecast performance at the peak. \textbf{A:} Simulated daily new infections for a two peak epidemic (dots) with five forecasts from both the ARIMA model (green lines) and the epimodulated ARIMA model (pink lines). \textbf{B:} Cross-validated estimates of the primary epimodulation parameter, $\hat{\theta}$. Note, $\hat{\theta}$ is estimated at zero until the initial epidemic growth begins to slow.}
    \label{fig:theta-concept}
\end{figure}

\section*{Results}

In a simple simulated epidemic with two waves, epimodulation allows the model to learn peak structure and improve performance (Figure \ref{fig:theta-concept}). Prior to the first epidemic wave the model estimates $\theta=0$; thus forecasts from the base ARIMA model and the epimodulated version are identical (Figure \ref{fig:theta-concept}A). As the first epidemic wave progresses, the model finds that larger values of $\theta$ improve retrospective performance of forecasts, arriving at a value of 1.01 by day 140 of the epidemic (Figure \ref{fig:theta-concept}B). During the second wave the peak-aware version more closely projects future data, particularly prior to the second peak (Figure \ref{fig:theta-concept}A).

\begin{figure}
    \centering
    \includegraphics[width=\textwidth]{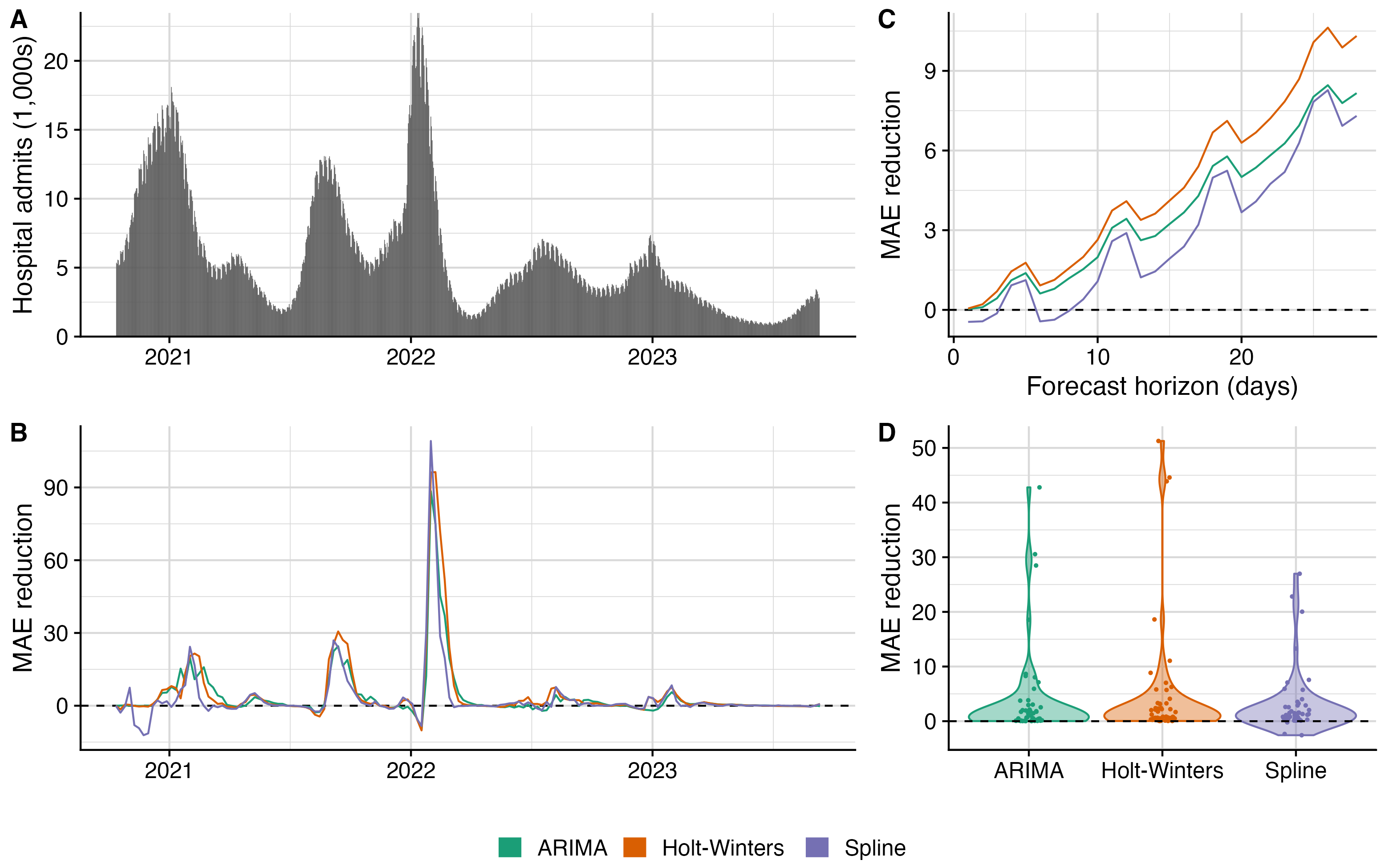}
    \caption{Impact of epimodulation on performance of COVID-19 hospital admissions forecasts from October 12, 2020 to September 11, 2023 for all 50 US states, Puerto Rico, Virgin Islands, and Guam. Mean absolute error (MAE) reduction measures the difference in forecasting error between an epimodulated model and its corresponding base model. \textbf{A:} Daily COVID-19 hospital admissions in the US provided by the COVIDHubUtils R package \cite{covidHubUtils}. Following COVID-19 Forecast Hub protocols, forecasts were made weekly at all horizons from 1 to 28 day(s) ahead \cite{covidforecasthub}. \textbf{B:} For each of the three methods, the average reduction in forecasting error across all locations and horizons for each forecast date during the study period (colored lines). \textbf{C:} For each of the three methods, the average reduction in forecasting error across all locations and dates for each forecast horizon. \textbf{D:} For each of the three methods, the average reduction in forecasting error across all dates and horizons for each location (points). Distributions across locations are summarized with violin plots. Positive MAE reduction values indicate that the epimodulation model outperformed the base forecast model, and the horizontal dashed line marks where the models performed similarly (Y=0). Absolute and percent improvement values are given in Table \ref{tab:summary-results}.
    \label{fig:covid-results}}
\end{figure}

Epimodulation substantially improves the performance of ARIMA, Holt-Winters, and Spline models in retrospective forecasts of COVID-19 hospital admissions from October 12, 2020 to September 11, 2023, a period spanning six epidemic waves in the US (Figure \ref{fig:covid-results}). However, the benefit is noticeable only around epidemic peaks (Figure \ref{fig:covid-results}B). Across the three models and all forecasts, we estimate an average MAE performance improvement of 10.2\% (range: 8.2-12.5\%); during the first Omicron wave (January 1-March 1, 2022), the average improvement is 20.7\% (range: 19.0-23.3\%) (Table \ref{tab:summary-results}). Improvements in model performance wane over the three years. The benefit of epimodulation increases with the length of the forecasting horizon, with an average MAE difference of 0.5 (range: -0.4-1.1) for 7-day ahead forecasts compared to 8.6 (range: 7.3-10.3) for 28-day ahead forecasts across the three models (Figure \ref{fig:covid-results}C, Table \ref{tab:summary-results}). Improvements are consistent across US regions with 100\%, 100\%, and 94.3\% of regions experiencing improvements using the epimodulated model for the ARIMA, Holt-Winters, and Spline models, respectively.

\begin{table}[]
\centering
\caption{Performance improvement of epimodulated versus base models, estimated by difference in mean absolute error (Absolute) and  percent reduction in mean absolute error (Percent). Results are presented for both influenza and COVID-19 across all forecast time periods and targets (Overall), for the peak time periods (Peak), for one- and four- week forecast horizons (one-week horizon and four-week horizon), and across all regions (Region). For both diseases we estimate peak performance during the largest epidemic wave during the time period. For COVID-19, the peak period is defined as January 1, 2022 to March 1, 2022 (Omicron); for influenza the peak period is defined as November 1, 2022 to March 1, 2023. Since COVID-19 forecasts were made on a daily timescale, we evaluate the 7 and 28 day ahead forecasts for the one- and four- week horizons, respectively.}
\label{tab:summary-results}
\resizebox{\textwidth}{!}{%
\begin{tabular}{cccccccccccc}
\hline
\multirow{2}{*}{Disease} &
  \multirow{2}{*}{Model} &
  \multicolumn{2}{c}{Overall} &
  \multicolumn{2}{c}{Peak} &
  \multicolumn{2}{c}{\begin{tabular}[c]{@{}c@{}}One week \\ horizon\end{tabular}} &
  \multicolumn{2}{c}{\begin{tabular}[c]{@{}c@{}}Four week \\ horizon\end{tabular}} &
  \multicolumn{2}{c}{Region} \\ \cline{3-12} 
                           &              & Absolute & Percent & Absolute & Percent & Absolute & Percent & Absolute & Percent & Absolute & Percent \\ \cline{1-2}
\multirow{3}{*}{COVID-19}  & ARIMA        & 3.8      & 10.0\%  & 28.9     & 19.0\%  & 0.8      & 3.9\%   & 8.2      & 12.3\%  & 3.8      & 5.0\%   \\
                           & Holt-Winters & 4.7      & 12.5\%  & 38.9     & 23.3\%  & 1.1      & 5.7\%   & 10.3     & 15.2\%  & 4.7      & 6.8\%   \\
                           & Spline       & 2.9      & 8.2\%   & 28.2     & 19.8\%  & -0.4     & -2.0\%  & 7.3      & 11.3\%  & 2.9      & 8.0\%   \\
\multirow{3}{*}{Influenza} & ARIMA        & 15.2     & 17.6\%  & 61.3     & 23.9\%  & 2.8      & 7.7\%   & 28.3     & 20.9\%  & 15.2     & 13.8\%  \\
                           & Holt-Winters & 15       & 17.7\%  & 61.2     & 24.4\%  & 3.5      & 9.6\%   & 28.1     & 21.2\%  & 15       & 16.8\%  \\
                           & Spline       & 33.1     & 23.2\%  & 104.7    & 28.2\%  & 14       & 16.0\%  & 52.4     & 26.5\%  & 33.1     & 26.9\%  \\ \cline{2-12} 
\end{tabular}%
}
\end{table}

\begin{figure}
    \centering
    \includegraphics[width=\textwidth]{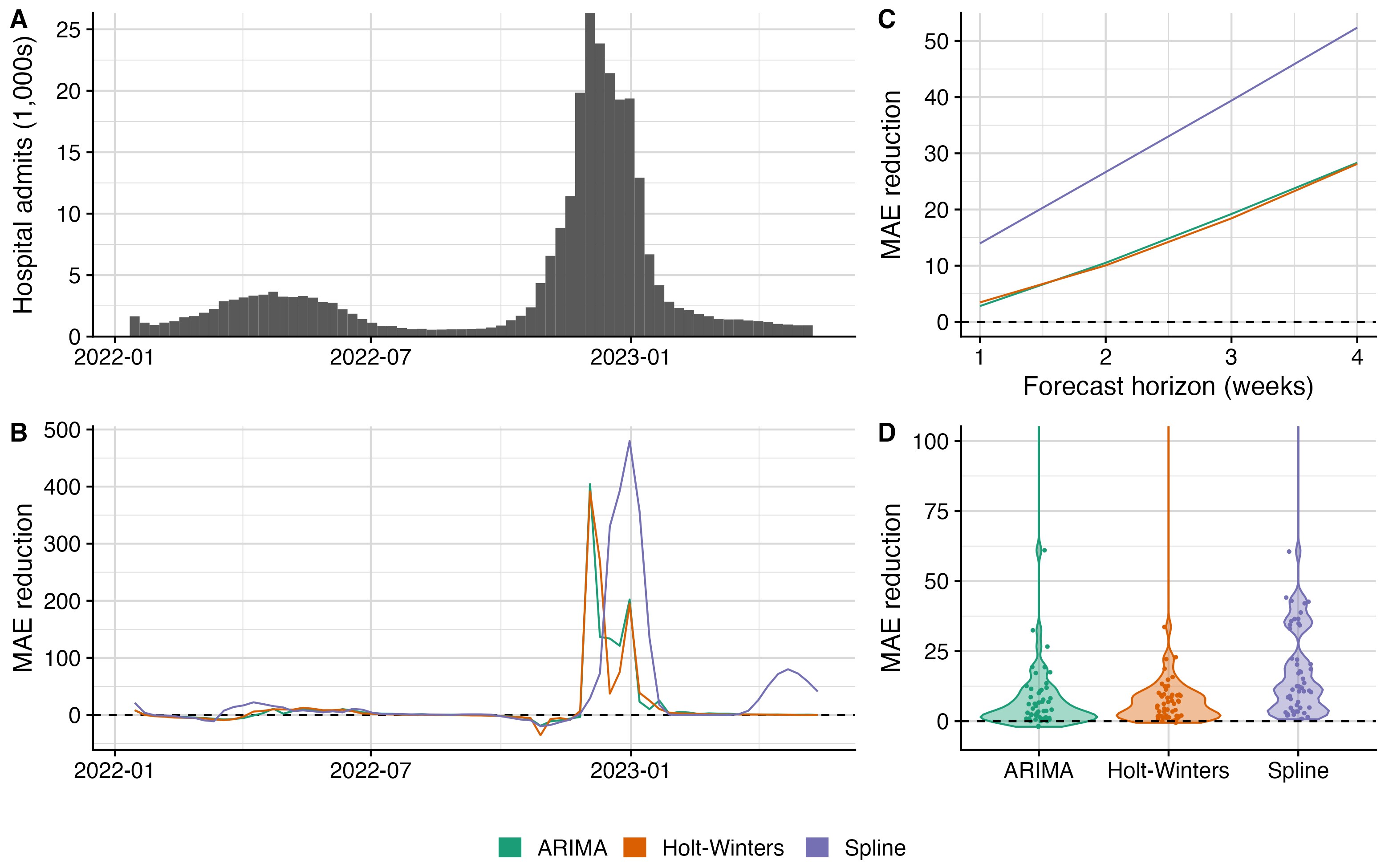}
    \caption{Impact of epimodulation on performance of influenza hospital admissions forecasts from January 10, 2022 to May 15, 2023 for all 50 US states, Puerto Rico, Virgin Islands, and Guam. Mean absolute error (MAE) reduction measures the difference in forecasting error between an epimodulated model and its corresponding base model. \textbf{A:} Daily influenza hospital admissions in the US provided by the COVIDHubUtils R package \cite{covidHubUtils}. Following FluSight Hub protocols, forecasts were made weekly at all horizons from 1 to 4 week(s) ahead \cite{flusight1}. \textbf{B:} For each of the three methods, the average reduction in forecasting error across all locations and horizons for each forecast date during the study period (colored lines). \textbf{C:} For each of the three methods, the average reduction in forecasting error across all locations and dates for each forecast horizon. \textbf{D:} For each of the three methods, the average reduction in forecasting error across all dates and horizons for each location (points). Distributions across locations are summarized with violin plots. Positive MAE reduction values indicate that the epimodulation model outperformed the base forecast model, and the horizontal dashed line marks where the models performed similarly (Y=0). Absolute and percent improvement values are given in Table \ref{tab:summary-results}.}
    \label{fig:flu-results}
\end{figure}

For influenza, forecast improvements are similarly concentrated around peaks, with a 19.5\% (range: 17.6-23.2\%) improvement overall compared to a 25.5\% (range: 23.9-28.2\%) improvement during the seasonal wave from November 1, 2022 to March 1, 2023 (Figure \ref{fig:flu-results}B). Again, the benefits of epimodulation increase with  forecast horizon (Table \ref{tab:summary-results} and Figure \ref{fig:flu-results}C) and are consistent across US regions (Figure \ref{fig:flu-results}D).

\begin{figure}
    \includegraphics[width=\linewidth]{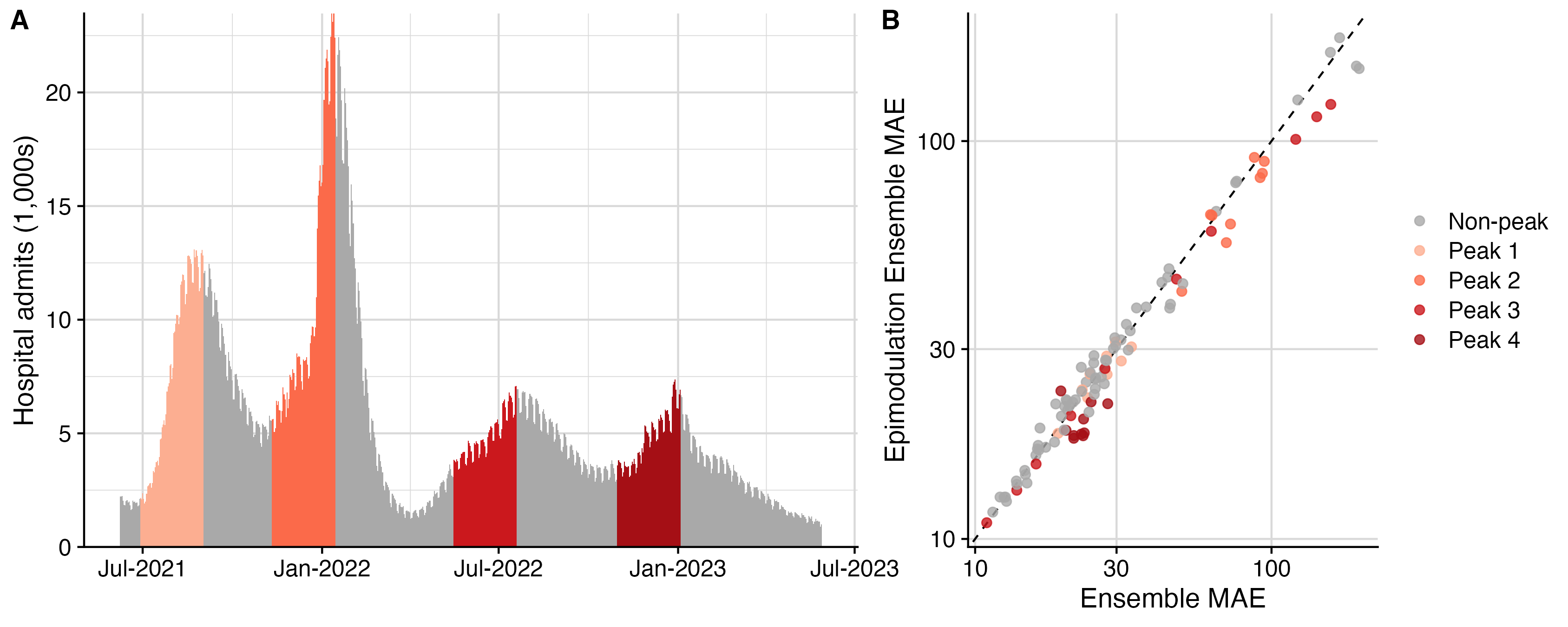}
  \caption{Epimodulation improves COVID-19 ensemble forecasts during  epidemiological peaks. \textbf{A}: Daily COVID-19 hospital admissions in the United States provided by the COVIDHubUtils R package \cite{covidHubUtils}. We applied the epimodulation algorithm to mean ensemble  1 to 28 day ahead forecast predictions made weekly from June 7, 2021 until May 29, 2023 and compared those with the original forecasts. Colored time periods indicate the nine weeks immediately before the four major peaks during that period. \textbf{B}: Epimodulation ensemble forecasts outperform the base ensemble forecasts particularly during peak time periods. Comparison between the weekly mean absolute error (MAE) of the ensemble and the epimodulated ensemble. Each dot represents the mean absolute error of the forecast across all fifty states and all forecast horizons, and dots are colored according to their time period from panel A. The dashed line indicates equal performance between the model forecasts for a specific date and points below the line indicate that the epimodulated ensemble forecasts outperformed the base ensemble forecast for that date. Note that both the x-axis and y-axis are on a log-scale.}
  \label{fig:date-improvement}
\end{figure}

We highlight the flexibility and impact of the epimodulation methodology by applying it to the gold standard COVID-19 ensemble forecast model (``COVIDhub-4-week-ensemble") produced by the COVID-19 Forecast Hub across all 50 states from June 7, 2021 until May 29, 2023 (Figure \ref{fig:date-improvement}A) \cite{covidhub,cdc_flusight}. We produced epimodulated ensemble forecasts during this time period and compared them to the base ensemble forecasts. Overall, we find that epimodulation improves the ensemble forecasts by 5.2\%, but we find that it improves the forecasts for the period before and during the epidemiological peaks by 10.0\%. In comparing the weekly forecasts, one can see that epimodulation mostly has a neutral impact, except for the improvements during the peak time periods (Figure \ref{fig:date-improvement}B).

\subsection*{Discussion}
Epimodulation enhances outbreak forecasts by incorporating the structure of the epidemic curve and anticipating the peak time from recent data. We tested it on a variety of forecasting methods, from basic empirical models to ensembles that combine multiple projections submitted to national forecasting hubs, for both COVID-19 and influenza. Epimodulation consistently improved accuracy across all tested models and diseases, with the greatest gains during epidemiological peaks and at longer forecast horizons (Table \ref{tab:summary-results}, Figure \ref{fig:covid-results}-\ref{fig:flu-results}). Although a 5-20\% performance boost to the COVID-19 Forecasting Hub ensemble may seem modest, it nonetheless provides a straightforward, validated way to enhance what many consider the gold standard in outbreak forecasting \cite{reich_collaborative_2022,reich_collaborative_2019,biggerstaff_improving_2022,biggerstaff_results_2018}. While we have not tested all possible models, these results suggest that epimodulation can be flexibly applied to improve forecast accuracy during critical periods without compromising overall performance.

The concept behind epimodulation originated early in the COVID-19 pandemic, when our group was developing forecast models to capture the unprecedented epidemiological dynamics unfolding in real time \cite{cramer_evaluation_2022,fox_real-time_2022,woody_projections_2020}. During this period, many models struggled because of uncertainties in behavioral responses, policy interventions, and the seasonal nature of COVID-19 transmission \cite{baker-seasonality,nixon_real-time_2022,lopez-case-forecasting}. In response, we pursued two parallel modeling strategies: (1) mechanistic models tailored to specific localities, augmented with real-time behavioral data to identify epidemiological change points \cite{woody_projections_2020,fox_real-time_2022}, and (2) flexible empirical models using machine learning to rapidly adapt to unexpected changes. However, across multiple waves, empirical forecasts consistently overpredicted outbreak peaks, lacking a built-in mechanism to anticipate the inevitable decline. A preliminary version of epimodulation addressed this shortcoming, significantly improving the forecast accuracy for both influenza and COVID-19 (Table \ref{tab:rt-wis}). The framework presented here is a generalized, flexible, and refined version of that initial approach.

Epimodulation was designed to improve forecast performance around epidemic peaks, and exceeded expectations in two key ways. First, it adapted in real time to varying wave shapes and seasonality, boosting accuracy even during the various waves of the COVID-19 pandemic (Figure \ref{fig:covid-results}). Second, these gains did not reduce the forecast performance for any specific region, time period or prediction horizon (Figures \ref{fig:covid-results} and \ref{fig:flu-results}). Thus, our cross-validation approach for estimating $\theta$ appears to be robust across different methods and epidemiological conditions. 

Empirical and mechanistic forecast models have both demonstrated strong performance in past forecast challenges, although empirical models have generally outperformed mechanistic ones \cite{reich_collaborative_2022,reich_collaborative_2019,biggerstaff_improving_2022,biggerstaff_results_2018}. However, both types of models struggle to accurately predict epidemiological peaks, which are shaped by a complex interplay of factors--including pathogen characteristics, changes in population-wide susceptibility, and the timing and effectiveness of pharmaceutical and non-pharmaceutical interventions \cite{bertozzi2020challenges}. Empirical models may fall short because historical patterns do not fully capture the range of possible future epidemic dynamics. Mechanistic models, on the other hand, may be limited by challenges in parameter estimation or by missing key elements of transmission dynamics. Epimodulation is a hybrid solution designed to address these challenges. It begins with empirical forecasts and applies a bias correction strategy inspired by epidemiological principles. In our implementation, the method corrects for the common tendency of forecasts to overpredict the epidemic peak, an error with multiple potential causes that are often difficult to isolate. In this sense, epimodulation functions as a constrained bias correction, where the adjustment is bounded between zero and one \cite{kennedy2001_bc}. Notably, similar bias correction methods have significantly improved the accuracy of climate and weather forecasts \cite{raftery2005using}. 

Although epimodulation shows promise, it has several limitations. First, it relies on data from previous epidemic waves, which may not be available for newly emerging threats (e.g., the initial months of COVID-19 or the 2022 international Mpox outbreak \cite{cramer_evaluation_2022, cdc_monkeypox_2022}). Second, the cross-validation process for estimating $\theta$ currently averages results across the entire estimation period; focusing on specific phases of previous waves (e.g., growth, peak, or decline) could be more effective. Third, in some cases, encoding specific mechanisms of transmission may outperform epimodulation. For instance, a Seasonal Auto-Regressive Integrated Moving Average (SARIMA) model might better capture simple seasonal drivers than an epimodulated ARIMA model. However, when the factors driving wave dynamics are complex, unknown, or unpredictable, epimodulation can offer a straightforward way to enhance performance. We caution that its impact should be retrospectively tested before using it in real-time outbreak forecasting.

Epimodulation offers a flexible framework for embedding epidemiological principles into a broad range of empirical forecast models, thereby providing a simple bridge between purely statistical and mechanistic approaches \cite{reich_collaborative_2019, biggerstaff_results_2016}. Although our implementation specifically augments models to capture population-wide declines in susceptibility, other predictable forces––such as seasonality or pathogen evolution––can be encoded if sufficient data are available to enable robust estimation. As with weather forecasting, we anticipate that incremental gains in forecast accuracy through methods like epimodulation, if widely tested and adopted, will yield substantial improvements in outbreak forecasting over decades \cite{bauer_quiet_2015}.

\newpage
\bibliographystyle{unsrt}
\bibliography{bib}

\section*{Appendix A1}
The epimodulation procedure originated during the early months of the COVID-19 pandemic, when we noticed that our empirical forecast models were overpredicting epidemiological peaks. Here we present the results from those initial experiments alongside their methodological explanations. We compare the performance of these models with the hub provided models including the baseline model that serves as a performance reference point and the ensemble model that produces forecasts from all contributed forecast models and has historically been the top performing forecast model \cite{cramer_evaluation_2022,flusight2,covidforecasthub,flusight1}. We contributed a version of the base Spline model to the FluSight Forecast hub from January 10, 2022 to June 20, 2022. After multiple waves, we found that the Base Spline model underperformed against the naive reference forecast model (Table \ref{tab:rt-wis}) with poor performance during epidemiological peaks where it tended to overpredict subsequent dynamics. We designed an early version of the epimodulation procedure that did not estimate $\theta$ using cross-validation and submitted that to the COVID-19 forecast hub from February 28, 2022 to February 26, 2024. Based on the promising results against the naive reference model (Table \ref{tab:rt-wis}), we submitted the same model to the FluSight Forecast hub from October 17, 2022 to May 15, 2023 \cite{flusight1,covidforecasthub}. We found that both models were able to outcompete the naive reference model during their respective time periods, though neither outperformed the gold standard ensemble forecast (Table \ref{tab:rt-wis}). Code for the models can be found at: https://github.com/gcgibson/semimech. Through retrospective analyses, we identified that the $\theta$ cross-validation procedure presented in the main manuscript was able to further improve forecast performance.

\setcounter{table}{0}
\renewcommand{\thetable}{A\arabic{table}}

\begin{table}[]
\centering
\caption{Real-time forecast performance of the base and epimodulated Spline model for COVID-19 and Influenza hospital admissions. Scores are compared with a Naive forecast model that is used as a reference for all forecast models and an ensemble forecast model that is the current gold standard for forecast performance. Performance measured as the average weighted interval score (WIS) across all forecast targets (Forecast target count) for each of the model versions. Lower WIS indicates better forecast performance. }
\label{tab:rt-wis}
\begin{tabular}{@{}cccccc@{}}
\toprule
 &
   &
  \begin{tabular}[c]{@{}c@{}}Forecast target\\ count\end{tabular} &
  \begin{tabular}[c]{@{}c@{}}Naive \\ WIS\end{tabular} &
  \begin{tabular}[c]{@{}c@{}}Model \\ WIS\end{tabular} &
  \begin{tabular}[c]{@{}c@{}}Ensemble \\ WIS\end{tabular} \\ \midrule
\multirow{2}{*}{Influenza} & Baseline Spline      & 5,060   & 22.3  & 30    & 19.6 \\
                           & epimodulated Spline & 6,240   & 123.1 & 117.4 & 90.2 \\
COVID-19                   & epimodulated Spline & 133,655 & 24.8  & 19.6  & 14.6 \\ \bottomrule
\end{tabular}%
\end{table}

\end{document}